\documentclass[a4paper,12pt]{article}

\usepackage{graphicx,amsmath,amsfonts,amssymb,cite}
\usepackage[dvips]{color}

\newcommand{\mj}{m_j}
\newcommand{\az}{\alpha Z}
\newcommand{\me}{m_e}
\newcommand{\mpr}{m_p}
\newcommand{\mf}{M_F}
\newcommand{\mi}{M_I}
\newcommand{\ga}{\gamma}
\newcommand{\ka}{\kappa}
\newcommand{\eps}{\epsilon}
\newcommand{\mub}{\mu_0}
\newcommand{\mun}{\mu_N}
\newcommand{\la}{\langle}
\newcommand{\ra}{\rangle}
\newcommand{\hfs}{hyperfine structure}

\newcommand{\vecb}{\vec{B}}

\newcommand{\De}{\Delta}

\newcommand{\dhfs}{\Delta E_{\rm HFS}}
\newcommand{\git}{g_I^{\prime}}
\newcommand{\Ft}{F^{\prime}}
\begin{document}
\begin{center}
\textbf{\large{ Zeeman effect of the hyperfine structure
levels in hydrogenlike ions }}\\ \vskip 1cm{} D.~L.~Moskovkin
and
 V.~M.~Shabaev,
 \\ \vskip 0.1cm{}
%\begin{footnotesize}
\emph{ Department of Physics, St. Petersburg State
University,\\ Oulianovskaya 1, Petrodvorets, St. Petersburg
198504, Russia}
%\end{footnotesize}
\end{center}
\vskip 0.5cm{}

\begin{abstract}
The fully relativistic theory of the Zeeman splitting of the $1s$
hyperfine structure levels in hydrogenlike ions is considered for
the magnetic field magnitude in the range from 1 to 10 T.
The second-order corrections to the Breit -- Rabi formula are
calculated and discussed. The results can be used for a precise
determination of nuclear magnetic moments from $g$ factor
experiments.
\end{abstract}

\section{Introduction}
Recent measurements of the $g$ factor of hydrogenlike carbon and
oxygen have reached an accuracy of about $7\cdot 10^{-10}$
\cite{her00,hae00,ver04}. Extensions of these measurements to ions
with nonzero nuclear spin $I$ would provide the basis for new
determinations of the nuclear magnetic moments \cite{wer01} and
the hyperfine structure (HFS) splitting in hydrogenlike ions.
Experimental investigations in this direction are anticipated in
the near future at GSI \cite{qui01}. Corresponding values of the
magnetic field are supposed to be in a range from 1 to 10 T.

For heavy ions with nonzero nuclear spin the ground state Zeeman
splitting caused by the magnetic field in the range under
consideration is much smaller than the hyperfine
splitting and, therefore, the consideration can be
conveniently reduced to the g factor value \cite{mosk04}. However,
for ions with the nuclear charge number $Z$ in the range $Z=1-20$,
which are being under current experimental investigations at
Mainz University, the Zeeman splitting is comparable with the
hyperfine splitting. This requires constructing the perturbation
theory for quasidegenerate states. To a good accuracy, the
solution of the problem is given by the well-known Breit -- Rabi
formula \cite{BR31, HKopf, bet57}. The aforesaid experimental
precision  has, however, shown the need for an improvement
of the Breit -- Rabi formula.

In the present paper, we improve the Breit -- Rabi formula for the
1s hyperfine structure levels by calculating the second-order
correction caused by the hyperfine interaction and the interaction
with the external magnetic field. The obtained results are
especially important for ions with $Z \leq 20$, where the 1s HFS
splitting can be comparable with the Zeeman splitting  if the
magnitude of the homogeneous magnetic field does not exceed 10 T.

Relativistic and Heaviside charge units ($\hbar=c=1~,\;
\alpha=e^{2}/4\pi$) are used in the paper, the charge of the
electron is taken to be $e<0$. In some important cases, the final
formulas contain $\hbar$ and $c$ explicitly to be applicable for
arbitrary system of units.

\section{The Breit -- Rabi formula}
We consider a H-like ion with nonzero nuclear spin $I$ in a state
with the total electron angular momentum $j=1/2$. The ion is
placed in a homogeneous magnetic field $\vec{B}$ directed along
the $z$ axis. The magnetic splitting is linear with respect to
$\vecb$ only if one of the following conditions is fulfilled:
either $\De E_{\rm mag}\ll \dhfs$ or $\De E_{\rm mag}\gg \dhfs$,
where $\dhfs=E(F+1)-E(F),\ E(F)=E_{n\ka} +\varepsilon_{\rm
hfs}(F)$, $F=I\pm 1/2$ is the total atomic angular momentum, and
$\varepsilon_{\rm hfs}(F)$ is the \hfs\ shift from the Dirac state
with the energy
\begin{equation}\label{en}
 E_{n\ka}=\frac{\ga+n_r}{N}\me\,.
\end{equation}
Here $n$ is the principal quantum number,
$\ka=(-1)^{j+l+\frac{1}{2}}(j+\frac{1}{2})$, $l=j\pm \frac{1}{2}$
defines the parity of the state, $n_r=n-|\ka|$ is the radial
quantum number, $\ga=\sqrt{\ka^2-(\az)^2}$,
$N=\sqrt{n_r^2+2n_r\ga+\ka^2}$, and $m_e$ is the electron mass. It
should be emphasized that in case the second inequality is
fulfilled $\De E_{\rm mag}$ must be much less than the
distance to other Dirac's levels. In the intermediate $\vecb$
case, $\De E_{\rm mag}\sim \dhfs$, we must take into account
mixing the HFS sublevels with the same $\mf$, where $\mf=-F, -F+1,
..., F-1, F$ is the $z$ projection of the total angular momentum. 
For the states with
$j=1/2$, there are only two HFS levels $F=I-1/2$ and $\Ft=I+1/2$
with the same $\mf=-I+1/2, ...,I-1/2$. This greatly simplifies the
theory. Denoting \footnote{In the present paper, the energy of a
Zeeman sublevel $\De E_{\rm mag}$ is counted with respect to the
mean energy of the \hfs\ doublet \cite{bet57, zap79}. 
If it is necessary to count $\De
E_{\rm mag}^{\rm cg}$ from the center of gravity of the HFS
doublet \cite{BR31,lB97}, one should use the relation
\begin{equation}
\De E_{\rm mag}^{\rm cg}= \De E_{\rm mag}-\frac{\dhfs}{2(2I+1)}\,.
\notag
\end{equation}
} $\De E_{\rm mag}= E-\frac{E(F)+E(F+1)}{2}$, one can derive for
the Zeeman splitting
%%%\small
\begin{equation}\label{BR1}
\De E_{\rm mag}(x)=\dhfs\left[a_1\mf x
        \pm
\frac{1}{2} \sqrt{1 +\frac{4\mf}{2I+1}c_1 x +c_2 x^2 }\,\,\right]\,,
\end{equation}
%%%\normalsize
where $x=\mub B/\dhfs$, $\mub=|e|\hbar/(2m_e c)$ is the Bohr
magneton,
\begin{align}\label{smallcoeff}
a_1&=-\git \,,   \\
c_1&= g_j +\git\,, \\
c_2&=(g_j +\git)^2\,,
\end{align}
$g_j $ is the bound-electron $g$ factor,
\begin{equation}\label{gj}
g_j = g_{\rm D}+\Delta g_{\rm QED}+\Delta g_{\rm rec}^{(e)}+
\Delta g_{\rm NS}+\Delta g_{\rm NP}\,,
\end{equation}
$g_{\rm D} $ is the bound-electron $g$ factor derived from the
Dirac equation \cite{zap79},
\begin{eqnarray}
g_{\rm D}=\frac{\ka}{j(j+1)}\left(\ka\frac{E_{n\kappa}}
{\me}-\frac{1}{2}\right)\,,
\end{eqnarray}
$\Delta g_{\rm QED}$ is the QED correction, $\Delta g_{\rm
rec}^{(e)}$ is the nuclear recoil correction to the bound-electron
$g$ factor, $\Delta g_{\rm NS}$ is the nuclear size correction,
$\Delta g_{\rm NP}$ is the nuclear polarization correction, $\git$
is the nuclear $g$ factor expressed in the Bohr magnetons,
\begin{equation}\label{defgit}
\git=\frac{\me}{\mpr}(g_I+\Delta g_{\rm rec}^{(n)})\,,
\end{equation}
$\mpr$ is the proton mass, $g_I=\mu/(\mun I)$ , $\mu =\langle
II|\mu{}_{z}|II\rangle$ is the nuclear magnetic moment, $\mu_z$ is
the $z$ projection of the nuclear magnetic moment operator
$\vec{\mu}$ acting in the space of nuclear wave functions
$|I\mi\rangle$ with the total angular momentum $I$ and its
projection $\mi$, $\mun=|e|\hbar/(2\mpr c)$ is the nuclear
magneton, and $\Delta g_{\rm rec}^{(n)}$ is the recoil correction
to the bound-nucleus $g$ factor (see section \ref{Numer}). Eq.
(\ref{BR1}) is usually called the Breit -- Rabi formula (see,
e.g., Refs. \cite{BR31, bet57, lB97, zap79}). It covers Zeeman
splitting from weak ($x\ll 1$) to strong ($x\gg 1$) fields
including the intermediate region. For $\Ft=I+\frac{1}{2}\,,\
\mf=\pm(I+\frac{1}{2})$ the splitting is linear in the first order
of perturbation theory under arbitrary magnetic induction,
\begin{equation}\label{BRl1}
\De E_{\rm mag}(x)=\dhfs\left[\frac{1}{2}
\pm d_1 x\right] \,,
\end{equation}
where
\begin{equation}
d_1 = \frac{1}{2}g_j-I\git\,
\end{equation}
and the ``$-$'' and ``$+$'' signs
refer to $\mf=-(I+\frac{1}{2})$ and
$\mf=I+\frac{1}{2}$, respectively.

For H-like ions with $I=1/2$, $F=0$ and $\Ft=1$ and, therefore,
the two mixed sublevels have $\mf=0$. In this case the Breit --
Rabi formula takes the form
\begin{equation}\label{BR2} \De E_{\rm mag}(x)=
  \pm\frac{\dhfs}{2} \sqrt{1 +c_2 x^2 } \,,
\end{equation}
and for $\mf=\pm 1$ the effect is described by Eq. (\ref{BRl1})
with $d_1 = \frac{1}{2}(g_j-\git)$.

It should be noted that in the original paper \cite{BR31} the
lowest-order nonrelativistic expression $g_j=(j+1/2)/(l+1/2)$ was
used for the electronic g factor, and the corrections depending on
$\frac{\me}{\mpr}g_I$ were introduced later \cite{bet57}.

If the magnetic field is strong, $\Delta E_{\rm mag} \gg\dhfs$,
Eqs. (\ref{BR1}), (\ref{BRl1}), and (\ref{BR2}) convert into
formulas for the anomalous Zeeman effect of the state with
$j=1/2$. On the contrary, assuming that the energy level shift
(splitting) due to interaction with $\vec{B}$ is much smaller than
the hyperfine-structure splitting, $\Delta E_{\rm mag}\ll \dhfs$,
we can express the linear-dependent part of this shift in terms of
the atomic $g$ factor,
\begin{equation}\label{magsplit}
\Delta E_{\rm mag}=g\mub B\mf \,,
\end{equation}
where, to the lowest-order approximation (see, e.g., Refs.
\cite{bet57, lB97}),
\begin{equation}\label{gat}
g(F)= g_{\rm D}Y_{\rm el}(F)-
\frac{\me}{\mpr}g_I Y_{\rm nuc}^{(\mu)}(F)\,,
\end{equation}
\begin{eqnarray} \label{Yel}
Y_{\rm el}(F)=\frac{F(F+1)+3/4-I(I+1)}{2F(F+1)}=
\begin{cases}
  -\frac{1}{2I+1}   &\text{for $F=I-\frac{1}{2}$}\\
  \frac{1}{2I+1}        &\text{for $F=I+\frac{1}{2}$}
\end{cases}\,,
\end{eqnarray}
\begin{eqnarray} \label{Ynucdip}
Y_{\rm nuc}^{(\mu)}(F)=\frac{F(F+1)+I(I+1)-3/4}{2F(F+1)}=
\begin{cases}
  \frac{2(I+1)}{2I+1}   &\text{for $F=I-\frac{1}{2}$}\\
  \frac{2I}{2I+1}        &\text{for $F=I+\frac{1}{2}$}
\end{cases}\,.
\end{eqnarray}
The total $1s$ $g$ factor value for a hydrogenlike ion with
nonzero nuclear spin can be represented by
\begin{eqnarray}\label{gtot}
g_{\rm tot}(F)&=&(g_{\rm D}+\Delta g_{\rm QED}+\Delta g_{\rm rec}^{(e)}+
\Delta g_{\rm NS}+\Delta g_{\rm NP}) Y_{\rm el}(F)
 \nonumber \\
&&-\frac{\me}{\mpr}(g_I+\Delta g_{\rm rec}^{(n)}) Y_{\rm
nuc}^{(\mu)}(F)+\delta g_{\rm HFS}^{(1s)}(F) \,,
\end{eqnarray}
where the HFS correction $\delta g_{\rm HFS}^{(1s)}(F)= \delta
g_{\rm HFS(\mu)}^{(1s)}(F)+\delta g_{\rm HFS(Q)}^{(1s)}(F)$ was
calculated in Ref. \cite{mosk04}.

\section{Corrections to the Breit -- Rabi formula for the $1s$ state}
The hamiltonian of a hydrogenlike ion can be written as
\begin{equation}
H = H_0 + V\,,
\end{equation}
where $H_0$ is the Dirac hamiltonian and
\begin{equation}
V = V_{\rm HFS} + V_{\vecb}\,.
\end{equation}
The hyperfine interaction operator is given by the sum
\begin{equation}
V_{\rm HFS}=V_{\rm HFS}^{(\mu)}+V_{\rm HFS}^{(Q)}\,,
\end{equation}
where $V_{\rm HFS}^{(\mu)}$ and $V_{\rm HFS}^{(Q)}$ are the
magnetic-dipole and electric-quadrupole hyperfine-interaction
operators, respectively. In the point-dipole approximation,
\begin{equation}\label{FB}
V_{\rm
HFS}^{(\mu)}=\frac{|e|}{4\pi}\frac{(\vec{\alpha}\cdot[\vec{\mu}\times
\vec{r}])}{r^3}\,,
\end{equation}
and, in the point-quadrupole approximation,
\begin{equation}
V_{\rm HFS}^{(Q)} = -\alpha \sum_{m=-2}^{m=2}Q_{2m}
\eta_{2m}^*(\vec{n})\,.
\end{equation}
Here $ Q_{2m}=\sum_{i=1}^Zr_i^2C_{2m}(\vec{n}_i)$ is the operator
of the electric-quadrupole moment of the nucleus,
$\eta_{2m}=C_{2m}(\vec{n})/r^3 $ is an operator that acts on
electron variables, $ \vec{n}=\vec{r}/r$, $\vec{n}_i=
\vec{r}_i/r_i$, $\vec{r}$ is the position vector of the electron,
$\vec{r}_i$ is the position vector of i-th proton in the nucleus,
$C_{lm}= \sqrt{4\pi/(2l+1)}\,Y_{lm}$, and $Y_{lm}$ is a spherical
harmonic. It must be stressed that the electric-quadrupole
interaction should be taken into account only for ions with $I >
1/2$.

The interaction of the ion with the magnetic field is represented
as
\begin{equation}
V_{\vecb} = V_{\vecb}^{(e)} + V_{\vecb}^{(n)}\,.
\end{equation}
Here $V_{\vecb}^{(e)}$ describes the interaction of the electron
with the homogeneous magnetic field,
\begin{equation}\label{W}
V_{\vecb}^{(e)}=-e(\vec{\alpha}\cdot
\vec{A})=\frac{|e|}{2}(\vec{\alpha}\cdot [\vecb\times\vec{r}])\,,
\end{equation}
where the vector $\vec{\alpha}$ incorporates the Dirac $\alpha$
matrices, and
\begin{equation}\label{Wn}
V_{\vecb}^{(n)}=-(\vec{\mu}\cdot \vecb)
\end{equation}
describes the interaction of the nuclear magnetic moment
$\vec{\mu}$ with $\vecb$.

We assume that the Zeeman splitting $\De E_{\rm mag}$ of the $1s$ HFS
levels $F=I-1/2$ and $\Ft=I+1/2$ is much smaller than the distance
to other levels but is comparable with $\dhfs^{(1s)}$. The
unperturbed eigenstates form a two-dimensional subspace
$\Omega=\{|1^{(0)}\ra,\ |2^{(0)}\ra\}$, where $
|1^{(0)}\ra=|10\frac{1}{2}IF\mf\ra,\
|2^{(0)}\ra=|10\frac{1}{2}I\Ft\mf\ra$, and $|nljIF\mf\rangle $
denotes the atomic wave function that corresponds to given values
of $F$ and $\mf $,
\begin{equation}\label{ket}
|nljIF\mf\rangle=\sum_{\mj, \mi}C_{j\mj
I\mi}^{F\mf}|nlj\mj\rangle|I\mi\rangle.
\end{equation}
Here $C_{j\mj I\mi}^{F\mf}$ are the Clebsch-Gordan coefficients,
$|nlj\mj\rangle$ are the unperturbed electron wave functions,
which are four-component eigenvectors of the Dirac equation for
the Coulomb field, and $|I\mi\rangle$ are the nuclear wave
functions.

Employing the perturbation theory for degenerate states 
\cite{sha02c} and keeping the three lowest-order
terms only, we get the following equation for the perturbed
energies
\begin{equation}\label{secur}
\begin{vmatrix}
h_0(F)+h_1(F)B+h_2(F)B^2-E &
\tilde{h}_1(F,\Ft)B+\tilde{h}_2(F,\Ft)B^2   \\
\tilde{h}_1(\Ft,F)B+\tilde{h}_2(\Ft,F)B^2
&h_0(\Ft)+h_1(\Ft)B+h_2(\Ft)B^2-E
\end{vmatrix}
=0 \,.
\end{equation}
Here $F=I-\frac{1}{2}$, $\Ft=I+\frac{1}{2}$,
\begin{equation}
h_0(k) = E(k)
\end{equation}
is the energy of the HFS level,
\begin{align}
h_1(k) &= \frac{1}{B}\biggl(\la k|V_{\vecb}|k\ra+2\sum_i^{(E_i\neq
E_{1,-1})}\frac{\la k|V_{\vecb}|i\ra\la i|V_{\rm HFS}|k\ra}
{E_{1,-1}-E_i}\biggr) \notag   \\ &+ (\Delta g_{\rm QED}+\Delta
g_{\rm rec}^{(e)}+ \Delta g_{\rm NS}+\Delta g_{\rm NP}) Y_{\rm
el}(k)\mub \mf -
 \Delta g_{\rm rec}^{(n)}Y_{\rm nuc}^{(\mu)}(k)\mun \mf \\
&= g_{\rm tot}(k)\mub \mf\,, \notag
\end{align}
\begin{equation}\label{h2k}
h_2(k)=\frac{1}{B^2}\sum_i^{(E_i\neq E_{1,-1})}\frac{|\la
k|V_{\vecb}|i\ra|^2}{E_{1,-1}-E_i}\,,
\end{equation}
\begin{align}\label{h1tjk}
\tilde{h}_1(j,k)&=\frac{1}{B}\biggl(\la j|V_{\vecb}|k\ra+
\sum_i^{(E_i\neq E_{1,-1})} \frac{\la j|V_{\vecb}|i\ra\la i|V_{\rm
HFS}|k\ra+\la j|V_{\rm HFS}|i\ra\la
i|V_{\vecb}|k\ra}{E_{1,-1}-E_i}\biggr)\notag \\ &+(\Delta _{\rm
QED}+\Delta _{\rm rec}+ \Delta _{\rm NS}+\Delta _{\rm NP})\mub \,,
\end{align}
\begin{equation}\label{h2tjk}
\tilde{h}_2(j,k)=\frac{1}{B^2}\sum_i^{(E_i\neq E_{1,-1})}\frac{\la
j|V_{\vecb}|i\ra\la i|V_{\vecb}|k\ra}{E_{1,-1}-E_i}\,,
\end{equation}
$j,\,k=F,\,\Ft$. $\Delta _{\rm QED}$, $\Delta _{\rm rec}$, $\Delta
_{\rm NS}$, and $\Delta _{\rm NP}$ are the QED, nuclear recoil,
nuclear size, and nuclear polarization corrections. They are
similar to the corresponding corrections to $ h_1(k)$ but have
a different angular factor. It should be noted that we have
neglected here terms
 describing virtual transitions into excited
nuclear states via the direct interaction of the nucleus with the
magnetic field. We assume that these terms are extremely small.
The calculation of $h_1(k)$ was discussed in detail in Ref.
\cite{mosk04}. Calculating the other matrix elements, we obtain
\begin{align}
h_2(k)B^2&=\frac{1}{(\az)^2}U(\az)(\mub B)^2/(\me c^2) \,,  \\
\tilde{h}_1(j,k)B&= \frac{1}{2}\frac{\sqrt{(I+1/2)^2-\mf^2}}{I+1/2}
\biggl[g_j+\git \notag  \\
&-\alpha^2 Z \frac{1}{3}\biggl\{ \git S(\az)+
(\az)^2\frac{11}{90}Q\left(\frac{\me c}{\hbar}\right)^2
\frac{2I+3}{2I}T(\az)\biggr\}\biggr]\mub B\,,   \\
\tilde{h}_2(j,k)&=0 \,.
\end{align}
Here
\begin{equation}\label{defU}
U(\az)=\frac{2}{9}(\az)^2[u_2(\az)+2u_{-1}(\az)]\,,
\end{equation}
\begin{equation}\label{defU-1}
u_{-1}(\az)=\me^3\tilde{R}_{-1}\,,
\end{equation}
\begin{equation}\label{defU2}
u_{2}(\az)=\me^3\tilde{R}_{2}\,,
\end{equation}
\begin{equation}\label{Rt1-1}
\tilde{R}_{-1}= \sum_{n}^{n\neq 1}\frac{1}{E_{1,-1}-E_{n,-1}}
\int_{0}^{\infty}(g_{1,-1}f_{n,-1}+ f_{1,-1}g_{n,-1})r^3\,
dr\int_{0}^{\infty}(g_{n,-1}f_{1,-1}+f_{n,-1}g_{1,-1})r^3\,dr\,,
\end{equation}
\begin{equation}\label{Rt12}
\tilde{R}_{2}=\sum_{n}\frac{1}{E_{1,-1}-E_{n,2}}
\int_{0}^{\infty}(g_{1,-1}f_{n,2}+ f_{1,-1}g_{n,2})r^3\,
dr\int_{0}^{\infty}(g_{n,2}f_{1,-1}+f_{n,2}g_{1,-1})r^3\,dr\,,
\end{equation}
\begin{eqnarray}\label{defS}
S(\alpha Z)=\frac{2}{3\az}(R_{2}+2R_{-1})\,,
\end{eqnarray}
\begin{equation}\label{R1-1}
R_{-1}= \sum_{n}^{n\neq 1}\frac{1}{E_{1,-1}-E_{n,-1}}
\int_{0}^{\infty}(g_{1,-1}f_{n,-1}+ f_{1,-1}g_{n,-1})r^3\,
dr\int_{0}^{\infty}(g_{n,-1}f_{1,-1}+f_{n,-1}g_{1,-1})\,dr\,,
\end{equation}
\begin{equation}\label{R12}
R_{2}=\sum_{n}\frac{1}{E_{1,-1}-E_{n,2}}
\int_{0}^{\infty}(g_{1,-1}f_{n,2}+ f_{1,-1}g_{n,2})r^3\,
dr\int_{0}^{\infty}(g_{n,2}f_{1,-1}+f_{n,2}g_{1,-1})\,dr\,,
\end{equation}
\begin{eqnarray}\label{defT}
T(\alpha Z)=-\frac{36}{11\me(\az)^3}\mathcal{R}_{2}\,,
\end{eqnarray}
\begin{equation}\label{otherR12}
\mathcal{R}_{2}= \sum_{n}\frac{1}{E_{1,-1}-E_{n,2}}
\int_{0}^{\infty}(g_{1,-1}f_{n,2}+ f_{1,-1}g_{n,2})r^3\,
dr\int_{0}^{\infty}(g_{n,2}g_{1,-1}+f_{n,2}f_{1,-1})\frac{1}{r}\,dr\,,
\end{equation}
$g_{n\ka}$ and $f_{n\ka}$ are the upper and lower radial
components of the Dirac wave function defined by
\begin{eqnarray}
\psi_{n\ka m}({\vec{r}})= \left(\begin{array}{c}
g_{n\ka}(r)\Omega_{\kappa m}({\vec{n}})\\
if_{n\ka}(r)\Omega_{-\kappa m}({\vec{n}})
\end{array}\right)\,,
\end{eqnarray}
and $Q=2\langle II|Q_{20}|II \rangle$ is the electric-quadrupole
moment of the nucleus. For the point-charge nucleus, the 
functions $S(\az)$ and $T(\az)$ are given by \cite{mosk04}
\begin{align}\label{s1}
S(\az)&=\frac{2}{3}\left\{\frac{2+\ga}{3(1+\ga)}+
\frac{2}{\ga(2\ga-1)}[1-\frac{\ga}{2}+(\az)^2]\right\}\notag \\
&=1+\frac{97}{36}(\az)^2+\frac{289}{72}(\az)^4+...\,
\end{align}
and
\begin{align}\label{t1}
T(\az)&=\frac{12(35+20\ga-32(\az)^2)}
{11\ga(1+\ga)^2[15-16(\az)^2]}\notag \\
 &=1+\frac{43}{33}(\az)^2 +\frac{2405}{1584}(\az)^4+... \,,
\end{align}
where $\ga=\sqrt{1-(\az)^2}$. Also the sum $\tilde{R}_{-1}$ can be
evaluated analytically, employing the method of generalized virial
relations for the Dirac equation \cite{sha91,sha03}. For the $1s$
state, applying formulas from Ref. \cite{gla02, mosk04}, we find
\begin{equation}\label{U-1}
u_{-1}(\az)=(\ga +1)\left[\frac{3}{4(\az)^2}-1\right]\,.
\end{equation}
Numerical evaluation of $u_{-1}(\az)$, $u_2(\az)$, $U(\az)$,
$S(\az)$, and $T(\az)$ for extended-charge nuclei is considered in
the next section.

Solving equation (\ref{secur}), we finally obtain for $\mf=-I+1/2,
...,I-1/2$
\begin{align}\label{BR3}
\De E_{\rm mag}(x) &= \dhfs^{(1s)}\biggl[a_1(1+\eps_1)\mf x +
 \eps_2 \frac{\dhfs^{(1s)}}{\me c^2}x^2 \notag \\
       & \pm
\frac{1}{2} \sqrt{1 +\frac{4\mf}{2I+1}c_1(1+\delta_1) x
+c_2(1+\delta_2+\mf^2\delta_3) x^2}\,\,\biggr]\,.
\end{align}
% where the ``$-$'' and ``$+$'' signs refer to the HFS states with
% $F$ and $\Ft=F+1$, respectively. 
Here
\begin{align}
\eps_1&=-\frac{1}{2\git}[\delta
g_{\rm HFS}^{(1s)}(F)+ \delta g_{\rm HFS}^{(1s)}(F+1)] \notag \\
       &= -\alpha^2 Z \frac{1}{3}\left[S(\az)-
 (\az)^2\frac{11Q}{30\git}\left(\frac{\me c}{\hbar}\right)^2
\frac{1}{I(2I-1)}T(\az)\right] \,,  \\
\eps_2&=\frac{1}{(\az)^2}U(\az)\,,
\\ \delta_1&=\frac{2I+1}{2(g_j+\git)}[\delta g_{\rm HFS}^{(1s)}(F+1)
- \delta g_{\rm HFS}^{(1s)}(F)] \notag \\
 &=-\alpha^2 Z \frac{1}{3(g_j+\git)}\left[\git S(\az)-
(\az)^2\frac{11}{90}Q\left(\frac{\me c}{\hbar}\right)^2
\frac{4I^2+4I+3}{I(2I-1)}T(\az)\right] \,,  \\
\delta_2&= -\alpha^2 Z \frac{2}{3(g_j+\git)}\left[\git S(\az)+
(\az)^2\frac{11}{90}Q\left(\frac{\me c}{\hbar}\right)^2
\frac{2I+3}{2I}T(\az)\right] \,,  \\
\delta_3&= \frac{1}{g_j+\git}\alpha^4 Z^3
\frac{22}{45}Q\left(\frac{\me c}{\hbar}\right)^2
\frac{1}{I(2I-1)}T(\az) \,.
\end{align}

For $\Ft=I+\frac{1}{2}\,,\ \mf=\pm(I+\frac{1}{2})$, in contrast to
Eq. (\ref{BRl1}), we have
\begin{equation}\label{BRnl}
\De E_{\rm mag}(x)=\dhfs^{(1s)}\biggl[\frac{1}{2} \pm
d_1(1+\eta_1) x+ \eta _2\frac{\dhfs^{(1s)}}{\me c^2} x^2\biggr]\,,
\end{equation}
where
\begin{align}
\eta_1 &=  \alpha^2 Z \frac{2}{3(g_j-2I\git)}\left[\git I S(\az)+
(\az)^2\frac{11}{90}Q\left(\frac{\me c}{\hbar}\right)^2
T(\az)\right] \,, \\ \eta _2 &=\eps_2=\frac{1}{(\az)^2} U(\az)\,,
\end{align}
and the ``$-$'' and ``$+$'' signs correspond to
$\mf=-(I+\frac{1}{2})$ and $\mf=I+\frac{1}{2}$, respectively.

If $I=1/2$, the electrical quadrupole interaction vanishes and one
can easily obtain for $\mf=0$:
\begin{equation}\label{BR4}
\De E_{\rm mag}(x)=\dhfs^{(1s)}\biggl[\eps_2
\frac{\dhfs^{(1s)}}{\me c^2}x^2
        \pm
\frac{1}{2} \sqrt{1 +c_2(1+\delta_2)x^2 }\,\biggr]
\end{equation}
with
\begin{equation}
\delta_2= -\frac{2\git}{3(g_j+\git)}\alpha^2 Z  S(\az)\,.
\end{equation}
For $I=1/2$, $\mf=\pm 1$, the effect is described by formula
(\ref{BRnl}) with
\begin{equation}
\eta_1 = \frac{\git}{3(g_j-\git)}\alpha^2 Z S(\az)\,.
\end{equation}

\section{Numerical results}\label{Numer}
In Table 1, we present the numerical results for the functions
$u_{-1}(\az)$, $u_2(\az)$, $U(\az)$, $S(\az)$, and $T(\az)$ (only
for the isotopes with $I>1/2$) defined by Eqs. (\ref{defU-1}),
(\ref{defU2}), (\ref{defU}), (\ref{defS}), and (\ref{defT}),
respectively. $u_{-1}^{\rm point}(\az)$, $S^{\rm point}(\az)$, and
$T^{\rm point}(\az)$ are the point-nucleus values obtained by
analytical formulas (\ref{U-1}), (\ref{s1}), and (\ref{t1}),
correspondingly. $u_{-1}^{\rm ext}(\az)$, $u_2^{\rm ext}(\az)$,
$U^{\rm ext}(\az)$, $S^{\rm ext}(\az)$, and $T^{\rm ext}(\az)$ are
the values calculated for the extended nuclear charge
distribution. The root-mean-square nuclear charge radii $\la
r^2\ra^{1/2}$ were taken from Ref. \cite{RnewAng}. 
The calculations were performed using the 
dual-kinetic-balance (DKB) basis set method \cite{DKB04}
with the
basis functions constructed from B-splines \cite{joh86,
joh88}. The uncertainties include the difference between the
results obtained with the Fermi and the homogeneously-charged
sphere model for the nuclear charge distribution as well as the
error arising from the uncertainty of $\la r^2\ra^{1/2}$.

In Table 2, we present the individual contributions to the $1s$
$g_j$ factor for some H-like ions with $I\neq 0$ in the range
$Z=1-20$.  The error ascribed to the Dirac point-nucleus value
results from the current uncertainty of the fine structure
constant, $1/\alpha=1/137.03599911(46)$ \cite{MoTaip}. The QED
correction includes the one-loop contribution to all orders in
$\alpha Z$ \cite{bei00,yer02b,lee05} and the existing  $\alpha Z$-expansion
QED terms of higher orders \cite{Pachtbp}.
 The recoil correction to the
bound-electron $g_j$ factor incorporates  the recoil effect of
first order in $m/M$, calculated to all orders in $\alpha Z$ in
\cite{sha01,sha02a}, and the existing $\alpha Z$-expansion terms
of orders $(m/M)^2$ and $\alpha (m/M)$ \cite{mar01}. The
nuclear-size correction was evaluated for the
homogeneously-charged-sphere model. The nuclear polarization
contribution to the ${1s}$ $g_j$ factor of light H-like ions
can be neglected \cite{nef02}.
The $g_j$ factor values 
given in Table 2 are used
for calculations of the coefficients in the Breit -- Rabi formula,
presented in Tables 3 and 4.

In Table 3, the numerical results for the coefficients in Eqs.
(\ref{BR1}), (\ref{BRl1}), (\ref{BR3}), and (\ref{BRnl}) are
listed  for some isotopes with $I\neq 1/2$ in the interval
$Z=1-20$. The numerical values of the coefficients in Eqs.
(\ref{BRl1}), (\ref{BR2}), (\ref{BRnl}), and (\ref{BR4}) for
$^{13}\rm{C}^{5+}$ ($I= 1/2$) are presented in Table 4. Since in
all the cases under consideration the absolute value of the recoil
correction to the bound-nucleus $g_I$ factor is smaller than
$10^{-11}$ \cite{mar01}, we actually have in Eq. (\ref{defgit}):
$\git=\frac{\me}{\mpr}g_I$.

\section{Discussion}

The energy separation between the ground-state HFS components
($F=I-1/2$ and $F^{\prime}=I+1/2$) of a hydrogenlike ion can be
written as \cite{ShHFS}
\begin{equation}\label{ShHFS}
\dhfs^{(1s)}=\frac{4}{3}\alpha (\az)^3\frac{\mu}{\mun}\frac{\me}{\mpr}
\frac{2I+1}{2I}\me c^2[A^{(1s)}(\az)(1-\delta^{(1s)}
(1-\epsilon^{(1s)})+x_{\rm rad}^{(1s)}]\,,
\end{equation}
where
\begin{equation}\label{A(1s)}
A^{(1s)}(\az)=\frac{1}{\ga(2\ga-1)}=1+\frac{3}{2}(\az)^2
+\frac{17}{8}(\az)^4+...
\end{equation}
is the relativistic factor, $\delta^{(1s)}$ is the nuclear charge
distribution correction, $\epsilon^{(1s)}$ is the nuclear
magnetization distribution correction (the Bohr -- Weisskopf
effect), and $x_{\rm rad}^{(1s)}$ is the QED correction.
Therefore, the dimensionless variable $x=\mub B/\dhfs^{(1s)}$ is
of order of $x_0\equiv \mub B/[\alpha (\az)^3\frac{\me}{\mpr}\me
c^2]$. Table 5 shows the value $x_0$ for various $B$ and $Z$. The
intervals of $B$ and $Z$, for which $x\sim 1$, are of special
interest (in the original paper \cite{BR31} the fields with
$0.1\leqslant x\leqslant 3$ were considered to be intermediate).

For the magnetic fields with the magnitude $B\sim 1-10 \, T$, that
are generally used in this kind of experiments, H-like ions with
$Z=1-20$ meet the requirement $x\sim 1$. For this reason, only
such ions are presented in Tables 2 -- 4.

For ions with $Z\leq 20$, the electrical quadrupole corrections
to the coefficients $a_1$, $c_1$, $c_2$, and $d_1$ are either by a
factor $10^{-3}-10^{-4}$ less than the magnetic dipole ones or
equal to zero, if $I=1/2$.

As one can see from Tables 3 and 4, the corrections $\epsilon_1$,
$\delta_1$, $\delta_2$, $\delta_3$, and $\eta_1$ provide  more precise
determinations of the coefficients in the Breit -- Rabi formula.

In the second-order approximation (\ref{secur}), formulas
(\ref{BR1}), (\ref{BR2}), (\ref{BR3}), and (\ref{BR4}) do not
contain $B$ to a power higher than two under the square root
(because of $h_2(F)=h_2(\Ft)$). For $B= 1-10 \, T$, an estimate of
the terms of higher orders with respect to $B$ indicates that the
contributions from these terms are negligibly small as compared
with both magnetic dipole and electrical quadrupole corrections.
However, it is very important to take into account $\epsilon_2
B^2$ and $\eta_2 B^2$ if $Z=1-20$. This is due to the fact that
these terms are comparable with the ones appearing from the
corrections to the Breit -- Rabi formula coefficients and the less
$Z$ is, the more appreciable the contributions from $\epsilon_2
B^2$ and $\eta_2 B^2$ become.

The Breit -- Rabi formula for the $1s$ state contains
$\dhfs^{(1s)}$, and the coefficients in the formula and
the corrections to them calculated above include the value of
$\mu/\mun$. Therefore, one can determine both $\dhfs^{(1s)}$ and
$\mu/\mun$ when carrying out the experiments on the Zeeman
splitting.

\section{Acknowledgements}
D.L.M. thanks N.~S.~Oreshkina for valuable advice when carrying
out the numerical calculations with the DKB basis set method.
This work was supported in part by
 RFBR (Grant No. 04-02-17574) and by INTAS-GSI
(Grant No. 03-54-3604).

\clearpage
\newpage

\begin{table}
\caption{The numerical results for the functions $u_{-1}(\az)$,
$u_2(\az)$, $U(\az)$, $S(\az)$, and $T(\az)$ (for the ions with
$I\neq 1/2$) defined by Eqs. (\ref{defU-1}), (\ref{defU2}),
(\ref{defU}), (\ref{defS}), and (\ref{defT}), accordingly.
$u_{-1}^{\text{point}}$, $S^{\text{point}}$, and
$T^{\text{point}}$ are the point-charge-nucleus values
obtained by formulas (\ref{U-1}), (\ref{s1}), and (\ref{t1}),
correspondingly. $u_{-1}^{\text{ext}} $, $u_{2}^{\text{ext}}$,
$U^{\text{ext}}$, $S^{\text{ext}}$, and $T^{\text{ext}}$ are the
 extended-charge-nucleus  values.  The values
of $\la r^2\ra^{1/2}$ are taken from Ref. \cite{RnewAng}. }
\begin{center}
\begin{tabular}{||c|l|l|l|l|l|l|l||}
\hline
Ion & $^{1}\rm{H}$& $^{13}\rm{C}^{5+}$ & $^{17}\rm{O}^{7+}$
 &$^{33}\rm{S}^{15+}$  & $^{43}\rm{Ca}^{19+}$
 &$^{53}\rm{Cr}^{23+}$  &$^{73}\rm{Ge}^{31+}$ \\ \hline
$Z$ &1 &6 &8 &16 &20 &24 &32    \\ \hline
$\la r^2\ra^{1/2}$, fm &0.879 &2.461 &2.695 &3.251
 &3.493 &3.659 &4.063 \\ \hline
$u_{-1}^{\text{point}} $ &28165.9 &780.079 &437.756
&107.663 &68.0544 &46.5408 &25.1555 \\ \hline
$u_{-1}^{\text{ext}} $ &28165.9 &780.079 &437.756
&107.663 &68.0548 &46.5412 &25.1561 \\ \hline
$u_{2}^{\text{ext}}$ &28167.0 &781.203 &438.880
&108.783 &69.1720  &47.6550 &26.2610 \\ \hline
$U^{\text{ext}}$ &0.999929 &0.997445 &0.995459 &0.981862
&0.971691 &0.959291 &0.927886 \\ \hline
$S^{\text{point}}$ &1.00014 &1.00518 &1.00923
&1.03749 &1.05927& 1.08659 &1.15986 \\ \hline
$S^{\text{ext}}$ &1.00014 &1.00518 &1.00922
&1.03737(1) &1.05901 & 1.08609(1) &1.15830(3) \\ \hline
$T^{\text{point}}$ &------------ &------------ &1.00446
&1.01805 &1.02846 &1.04145 &1.07586 \\ \hline
$T^{\text{ext}}$ &------------ &------------ &1.00357(2)
&1.01577(4) &1.0253(1) &1.0373(1) &1.0687(1) \\ \hline
\end{tabular}
\end{center}

\begin{center}
\begin{tabular}{||c|l|l|l|l|l||}
\hline
Ion & $^{129}\rm{Xe}^{53+}$ & $^{131}\rm{Xe}^{53+}$
 & $^{207}\rm{Pb}^{81+}$ & $^{209}\rm{Bi}^{82+}$
 &    $^{235}\rm{U}^{91+}$ \\ \hline
$Z$ &54 &54 &82 &83 &92    \\ \hline
$\la r^2\ra^{1/2}$, fm &4.776 &4.781 &5.494 &5.521 &5.829 \\ \hline
$u_{-1}^{\text{point}} $ &7.35001 & 7.35001
&1.97162 &1.87551 &1.15612\\ \hline
$u_{-1}^{\text{ext}}$ &7.35130(1) & 7.35130(1)
&1.97598(1) &1.88008(1) &1.16343(1) \\ \hline
$u_{2}^{\text{ext}}$ &8.41758 & 8.41758
&2.95636(1) &2.85645(1)  &2.09992(2)\\ \hline
$U^{\text{ext}}$ &0.797806(1) &0.797806(1) &0.549690(2)
&0.539399(2) &0.443386(4) \\ \hline
$S^{\text{point}}$  &1.54221 & 1.54221
&2.99051 &3.09142 &4.37922 \\ \hline
$S^{\text{ext}}$ &1.5249(2) &  1.5249(2)
&2.7193(14) &2.7907(16) &3.583(3) \\ \hline
$T^{\text{point}}$ &------------  &1.24668
&--------------- &1.82424 &2.20685 \\ \hline
$T^{\text{ext}}$ &------------  &1.2216(2)
&--------------- &1.6803(7) &1.933(1) \\ \hline
\end{tabular}
\end{center}
\end{table}

\begin{table}
\caption{  The individual contributions to the ${1s}$-electron $g_j$ factor of
hydrogenlike ions with nonzero nuclear spin and the nuclear
charge in the range $Z=1-20$.
The values of $\la r^2\ra^\frac{1}{2}$ are the same as in Table 1. }
\begin{center}
\begin{tabular}{||c|r@{.}l|r@{.}l|r@{.}l|r@{.}l||}
\hline
 Ion
&\multicolumn{2}{c|}{$^{13}\rm{C}^{5+}$}
&\multicolumn{2}{c|}{$^{17}\rm{O}^{7+}$}
&\multicolumn{2}{c|}{ $^{33}\rm{S}^{15+}$}
&\multicolumn{2}{c||}{$^{43}\rm{Ca}^{19+}$} \\ \hline

 $g_{\rm D}$&
 1&99872135439(1) &1&99772600306(2)      &1&99088058242(6)
 &1&9857232037(1)
\\ \hline

$\De g_{\rm{QED}}$
 &0&00232014777(3)  &0&00232089875(11)      &0&0023273918(32)
 &0&0023333328(100)
 \\ \hline

$\De g_{\rm{rec}}^{(e)}$&
 0&00000008087&0&00000011001
&0&00000022876 & 0&0000002761 \\ \hline

 $\De g_{\rm{NS}}$
 &0&00000000040  &0&00000000155(1)     &0&0000000386(12)
 &0&0000001141(1)
\\ \hline

   $g_j$&2&00104158344(3)
&2 &00004701337(11)
&1&993208242(3) &1&988056927(10) \\ \hline
\end{tabular}
\end{center}

\end{table}
%%%\end{document}

\begin{table}
\caption{The numerical values of the coefficients in Eqs. (\ref{BR1}),
(\ref{BRl1}), (\ref{BR3}), and (\ref{BRnl})
for H-like ions with $I \neq 1/2$ and $Z=1-20$.
The values of $\mu/\mun$ and $Q$ are taken from Refs. \cite{rag89}
and \cite{PePy}, respectively.}
\begin{center}
\begin{tabular}{||c|r@{.}l|r@{.}l|r@{.}l||}
\hline
 Ion
&\multicolumn{2}{c|}{$^{17}\rm{O}^{7+}$}
&\multicolumn{2}{c|}{ $^{33}\rm{S}^{15+}$}
&\multicolumn{2}{c||}{$^{43}\rm{Ca}^{19+}$} \\
\hline
$I$ &\multicolumn{2}{c|}{5/2}
&\multicolumn{2}{c|}{3/2}
&\multicolumn{2}{c||}{7/2}  \\
\hline
$\mu/\mun$
&\multicolumn{2}{c|}{-1.89379(9)}
&\multicolumn{2}{c|}{0.6438212(14)}
&\multicolumn{2}{c||}{-1.317643(7)} \\ \hline
$Q$, barn
&\multicolumn{2}{c|}{-0.02558(22)}
&\multicolumn{2}{c|}{-0.0678(13)}
&\multicolumn{2}{c||}{-0.0408(8)} \\ \hline
$a_1$ &0&00041256(2)   &-0&0002337573(5) &0&000205032(1) \\ \hline
$\epsilon_1$&-0&0001433 &-0&0002947
  &-0&0003759
 \\ \hline
$a_1(1+\epsilon_1)$ &0&00041250(2) &-0&0002336884(5) &0&000204955(1) \\ \hline
$\epsilon_2(=\eta_2)$
&\multicolumn{2}{c|}{292.087}
&\multicolumn{2}{c|}{72.0242}
&\multicolumn{2}{c||}{45.6181} \\ \hline
$c_1$ &1&99963441(2) &1&993442046(4)  &1&98785187(2)  \\ \hline
$\delta_1$  &0&00000002957 &-0&00000003461 &0&00000003874
 \\ \hline
$c_1(1+\delta_1)$ &1&99963447(2) &1&993441977(4) &1&98785194(2) \\ \hline
$c_2$ &3&99853778(8) &3&97381119(2) &3&95155504(6)  \\ \hline
$\delta_2$ &0&00000005914 &-0&00000006905
&0&00000007759  \\ \hline
$\delta_3$ &0&0 &-0&00000000004
&-0&00000000001  \\ \hline
$c_2(1+\delta_2)$ &3&99853802(8)&3&97381092(2) &3&95155535(6) \\ \hline
$c_2\delta_3$ &0&0 &-0&00000000017
&-0&00000000003  \\ \hline
$d_1$ &1&00105487(5) &0&996253509(3) &0&99474606(1) \\ \hline
$\eta_1$&-0&0000001477 &0&0000001037 &-0&0000002712  \\ \hline
$d_1(1+\eta_1)$ &1&00105473(5) &0&996253612(3)  &0&99474579(1) \\ \hline
\end{tabular}
\end{center}
\end{table}

%%%\newpage
%%%\clearpage

\begin{table}
\caption{The numerical values of the coefficients in Eqs. (\ref{BR2})
and (\ref{BR4}) for $^{13}\rm{C}^{5+}$ ($I= 1/2$).
$\mu/\mun$ is taken from \cite{rag89}. }
\begin{center}
\begin{tabular}{||c|r@{.}l||}
\hline
Ion &
\multicolumn{2}{c||}{$^{13}\rm{C}^{5+}$}   \\ \hline
$\mu/\mun$&
\multicolumn{2}{c||}{0.7024118(14)}     \\ \hline
$\epsilon_2(=\eta_2)$
&\multicolumn{2}{c||}{520.302}   \\ \hline
$c_2$ &4&007230231(6)    \\ \hline
$\delta_2$ &-0&00000008183   \\ \hline
$c_2(1+\delta_2)$ &4&007229903(6)  \\ \hline
$d_1$ &1&0001382800(8)   \\ \hline
$\eta_1$ &0&00000004095   \\ \hline
$d_1(1+\eta_1)$ &1&0001383209(8)    \\ \hline
\end{tabular}
\end{center}
\end{table}

\begin{table}
\caption{The values $x_0=\mub B/[\alpha (\az)^3\frac{\me}{\mpr}\me
c^2]$ for various $B$ and $Z$.}
\begin{center}
\begin{tabular}{||c|c|c|c|c|c|c|c||}
\hline $B$, T & 0.5 &  1 & 5 & 10 & 50 & 100  \\
  & & & & & & \\
$Z$ & & & & & & \\ \hline 1 &$3.7\cdot10$
 &$7.3\cdot10$ &$3.7\cdot10^{2}$ &$7.3\cdot10^{2}$ &
$3.7\cdot10^{3}$& $7.3\cdot10^{3}$ \\ \hline 4 &$5.7\cdot10^{-1}$
& 1.1 & 5.7 & $1.1\cdot10$ &$5.7\cdot10$ & $1.1\cdot10^{2}$
\\ \hline 7 &$1.1\cdot10^{-1}$ &$2.1\cdot10^{-1}$& 1.1 & 2.1
&$1.1\cdot10$ &$2.1\cdot10$  \\ \hline 15 &$1.1\cdot10^{-2}$
&$2.2\cdot10^{-2}$&$1.1\cdot10^{-1}$ &$2.2\cdot10^{-1}$ &1.1 & 2.2
\\ \hline 20 &$4.6\cdot10^{-3}$ &$9.2\cdot10^{-3}$ &$4.6\cdot10^{-2}$
 &$9.2\cdot10^{-2}$
&$4.6\cdot10^{-1}$ & $9.2\cdot10^{-1}$ \\ \hline 30
&$1.4\cdot10^{-3}$ &$2.7\cdot10^{-3}$ &$1.4\cdot10^{-2}$
&$2.7\cdot10^{-2}$ &$1.4\cdot10^{-1}$ &$2.7\cdot10^{-1}$
\\ \hline 80 &$7.2\cdot10^{-5}$ &$1.4\cdot10^{-4}$ &$7.2\cdot10^{-4}$
&$1.4\cdot10^{-3}$ &$7.2\cdot10^{-3}$ &$1.4\cdot10^{-2}$
\\ \hline
\end{tabular}
\end{center}
\end{table}


\begin{thebibliography}{99}


%%%\includegraphics[scale=1.80]{graph.eps}\\

\bibitem{her00}
N.~Hermanspahn, H.~H\"{a}ffner, H.-J.~Kluge, W.~Quint, S.~Stahl,
J.~Verd\'{u}, and G.~Werth, Phys. Rev. Lett. $\boldsymbol{84}$
427 (2000).
%
\bibitem{hae00}
H. H{\"a}ffner, T. Beier, N. Hermanspahn, H.-J. Kluge, W. Quint,
S. Stahl, J. Verd{\'u}, and G. Werth, Phys. Rev. Lett. {\bf 85},
5308 (2000).
%
\bibitem{ver04}
J.L. Verd{\'u}, S. Djeki{\'c}, S. Stahl, T. Valenzuela, M. Vogel, G.
Werth, T. Beier, H.-J. Kluge,  W. Quint, Phys. Rev. Lett. {\bf
92}, 093002 (2004).
%
%
\bibitem{wer01}
 G. Werth, H. H{\"a}ffner, N. Hermanspahn, H.-J. Kluge,
W. Quint, J. Verd{\'u}, in
 {\it The Hydrogen Atom}, edited by S.G. Karshenboim {\it et al.} (Springer,
 Berlin, 2001), p. 204.
%
%
\bibitem{qui01}
W. Quint, J. Dilling, S. Djekic, H. H{\"a}ffner,
 N. Hermanspahn, H.-J. Kluge, G. Marx, R. Moore, D. Rodriguez,
J. Sch{\"o}nfelder, G. Sikler, T. Valenzuela, J. Verd{\'u}, C.
Weber, and G. Werth,
 Hyperfine  Interactions, $\boldsymbol{132}$, 453 (2001).
%
%
\bibitem{mosk04}
D.L.~Moskovkin, N.S.~Oreshkina, V.M.~Shabaev, T.~Beier, G.
Plunien, W.~Quint, and G. Soff, Phys. Rev. A {\bf 70}, 032105
(2004).
%
%
\bibitem{BR31}
G. Breit and I.I. Rabi, Phys. Rev. {\bf 38}, 2082 (1931).
%
%
\bibitem{HKopf}
H. Kopfermann, \emph{Kernmomente, 2. Auflage}. Akademische
Verlagsgesellschaft, Frankfurt am Main, 1956.
%
\bibitem{bet57}
H.A. Bethe and E.E. Salpeter, \emph{Quantum mechanics of one and two
electron atoms}, Springer, Berlin, 1957.
%
%
\bibitem{zap79}
S.A. Zapryagaev, Opt. Spektrosk. {\bf 47}, 18 (1979) [Opt.
Spectrosc. $\boldsymbol{47}$, 9 (1979)].
%
\bibitem{lB97}
L.~Bergman and C.~Schaefer, \emph{Constituents of matter: atoms,
molecules, nuclei and particles}, de Gruyter, Berlin, 1997.
%
%
\bibitem{sha02c}
V.M. Shabaev, Phys. Rep. {\bf 356}, 119 (2002).
\bibitem{sha91}
V.M. Shabaev, J. Phys. B $\boldsymbol{24}$  4479 (1991).
%
\bibitem{sha03}
V.M. Shabaev, in {\it Precision Physics of Simple Atomic Systems},
edited by S.G. Karshenboim and V.B. Smirnov (Springer, Berlin,
2003), p. 97; E-print/physics/0211087 (2002).
%
%
\bibitem{gla02}
D.A. Glazov and V.M. Shabaev, Phys. Lett. A, {\bf 297}, 408
(2002).
%
%
\bibitem{RnewAng}
I. Angeli, At. Data Nucl. Data Tables $\boldsymbol{87}$, 185 (2004).
%
%
\bibitem{DKB04}
V.M. Shabaev, I.I. Tupitsyn, V.A. Yerokhin, G. Plunien, and G.
Soff, Phys. Rev. Lett. {\bf 93}, 130405 (2004).
%
\bibitem{joh86}
W.R. Johnson and J. Sapirstein, Phys. Rev. Lett. {\bf 57}, 1126
(1986).
%
\bibitem{joh88}
W.R. Johnson, S.A. Blundell, and J. Sapirstein, Phys. Rev. A.
$\boldsymbol{37}$, 307 (1988).
%
%
\bibitem{MoTaip}
P.J. Mohr and B.N. Taylor, Rev. Mod. Phys. $\boldsymbol{77}$, 1 (2005).
%
\bibitem{bei00}
T.~Beier, I.~Lindgren, H.~Persson, S.~Salomonson,
P.~Sunnergren, H.~H{\"a}ffner, and N.~Hermanspahn,
Phys. Rev. A {\bf 62}, 032510 (2000).
%
\bibitem{yer02b}
V.A.~Yerokhin, P.~Indelicato, and 
V.M.~Shabaev, Phys. Rev. Lett. {\bf 89},
 143001 (2002); Phys. Rev. A {\bf 69}, 052503 (2004).
%
\bibitem{lee05}
R.N. Lee, A.I. Milstein, I.S. Terekhov,
and S.G. Karshenboim, Phys. Rev. A {\bf 71}, 052501 (2005).
%
%
\bibitem{Pachtbp}
K. Pachucki, A. Czarnecki, U. D. Jentschura, and V.A. Yerokhin, 
Phys. Rev. A {\bf 72}, 022108 (2005).
%
\bibitem{sha01}
V.M. Shabaev, Phys. Rev. A {\bf 64}, 052104 (2001).
%
\bibitem{sha02a}
V.M. Shabaev and V.A. Yerokhin, Phys. Rev. Lett. {\bf 88}, 091801
(2002).
%
\bibitem{mar01}
A.P. Martynenko and R.N. Faustov, Zh. Eksp. Teor. Fiz. {\bf 120},
539 (2001) [JETP {\bf 93}, 471 (2001)].
%
\bibitem{nef02}
A.V.~Nefiodov, G.~Plunien, and G.~Soff,
Phys. Rev. Lett. {\bf 89}, 081802 (2002).
%
\bibitem{ShHFS}
V.M. Shabaev, M.B. Shabaeva, and I.I. Tupitsyn, Phys. Rev. A.
$\boldsymbol{52}$, 3686 (1995).
%
%
\bibitem{rag89}
P. Raghavan, At. Data Nucl. Data Tables, $\boldsymbol{42}$,
189 (1989).

\bibitem{PePy}
P. Pyykk{\"o}, Mol. Phys., 2001, Vol. {\bf 99}, No. 19, 1617.

%
%
\end{thebibliography}
\end{document}